\documentclass[11pt,a4paper]{article}
\usepackage[T1]{fontenc}
\usepackage{lmodern,amsmath,amssymb,amsthm,mathtools,booktabs,microtype}
\usepackage[margin=25mm]{geometry}
\usepackage{graphicx}
\usepackage[colorlinks=true,urlcolor=blue,citecolor=blue,linkcolor=blue]{hyperref}
\usepackage{enumitem}
\setlist{itemsep=2pt,topsep=4pt}
\newtheorem{theorem}{Theorem}[section]
\newtheorem{proposition}[theorem]{Proposition}
\newtheorem{lemma}[theorem]{Lemma}

\theoremstyle{remark}
\newcommand{\E}{\mathbb E}
\newcommand{\KL}{D_{\mathrm{KL}}}
\newcommand{\Var}{\operatorname{Var}}

\newcommand{\R}{\mathbb R}
\title{A proof of the Eskenazis--Nayar--Tkocz conjecture\ on entropy maximization for cube projections}
\hypersetup{
  pdftitle={A proof of the Eskenazis--Nayar--Tkocz conjecture on entropy maximization for cube projections},
  pdfauthor={Vu Khac Ky}
}
\author{Vu Khac Ky\\\small FPT University, Hanoi, Vietnam\\\small\texttt{kyvk2@fe.edu.vn}}
\date{}
\begin{document}
\maketitle
\begin{abstract}
We complete the Eskenazis--Nayar--Tkocz (ENT) conjecture. Among the
one-dimensional projections of the uniform measure on a cube, differential
entropy is maximized precisely in directions with equal coordinate
magnitudes. Pang proved the conjecture in every dimension at least $150$.
We prove the remaining dimensions using polynomial entropy bounds in
dimensions three through six and exact certificates for derivatives of
the logarithm of the diagonal projection density. The proof uses integer
arithmetic and certified one-dimensional integration. We also derive
quantitative stability estimates and determine the optimal asymptotic
coefficient of a linear stability bound. The source code and the outputs
of all finite verifications accompany the paper.
\end{abstract}
\noindent\textbf{Keywords.} Differential entropy; cube projections; coefficient balancing;
Schur concavity; computer-assisted proof.\\
\textbf{2020 Mathematics Subject Classification.} 60E15, 94A17, 52A40.
\par\smallskip
\noindent\textbf{Code.} \url{https://github.com/vukhacky/ent-conjecture/}

\section{Introduction}
Let $U_1,\ldots,U_n$ be independent random variables uniformly distributed
on $[-1,1]$. A one-dimensional orthogonal projection of the uniform
measure on the cube $[-1,1]^n$ has the distribution of
\[
 S_a=\sum_{i=1}^n a_iU_i,\qquad a\in\R^n,\qquad \sum_{i=1}^na_i^2=1.
\]
All these variables have mean zero and variance $1/3$. Their densities,
and hence their differential entropies, depend on the direction $a$.
Writing $p_a$ for the density of $S_a$, we use natural logarithms and set
\[
 h(S_a)=-\int_{\R}p_a(x)\log p_a(x)\,dx,
 \qquad T_n=\frac1{\sqrt n}\sum_{i=1}^n U_i.
\]
Each projection has a bounded, compactly supported density, so its entropy
is finite. The Eskenazis--Nayar--Tkocz (ENT) conjecture asks whether
distributing the squared coefficients equally maximizes this entropy.
Since the summands are symmetric, only the magnitudes of the
coefficients matter.

\begin{theorem}\label{thm:main}
For every integer $n\ge1$ and every $a\in\R^n$ with $\|a\|_2=1$,
\begin{equation}\label{eq:main}
 h(S_a)\le h(T_n).
\end{equation}
Equality holds if and only if $|a_i|=1/\sqrt n$ for every $i$.
\end{theorem}

The essential foundation for this result is Pang's seminal work
\cite{Pang}, which proves Theorem~\ref{thm:main} for every $n\ge150$,
including the equality statement. Our contribution to the conjecture
is the proof of the remaining dimensions $n\le149$. The finite
verification below also covers $n=150$, so there is overlap at the
boundary. We retain Pang's large-dimensional theorem as an explicit
input to the proof of Theorem~\ref{thm:main}.

The conjecture was posed in \cite[Question~9]{ENT2018}.
\subsection{Information, geometry and probability}
The fixed variance makes entropy a direct measure of departure from a
Gaussian distribution. Let $G$ have law $N(0,1/3)$. For every unit $a$,
\[
 D(p_{S_a}\Vert p_G)=h(G)-h(S_a).
\]
Thus \eqref{eq:main} says precisely that the diagonal projection
minimizes relative entropy to $G$ within this family. This is the sense
in which the diagonal is the most Gaussian direction at fixed $n$.

The comparison also complements the monotonicity theorem of Artstein,
Ball, Barthe and Naor \cite{ABBN2004}, which states that the entropy of
normalized sums of independent identically distributed square-integrable
random variables is nondecreasing with the number of summands. In our
setting this gives $h(T_n)\le h(T_{n+1})$, and the entropic central limit
theorem gives convergence to $h(G)$ \cite{Barron1986}.
The fixed-dimensional assertion identifies $T_n$ as an upper bound for
all choices of weights with the same number of summands.

\begin{figure}[t]
\centering
\includegraphics[width=\linewidth]{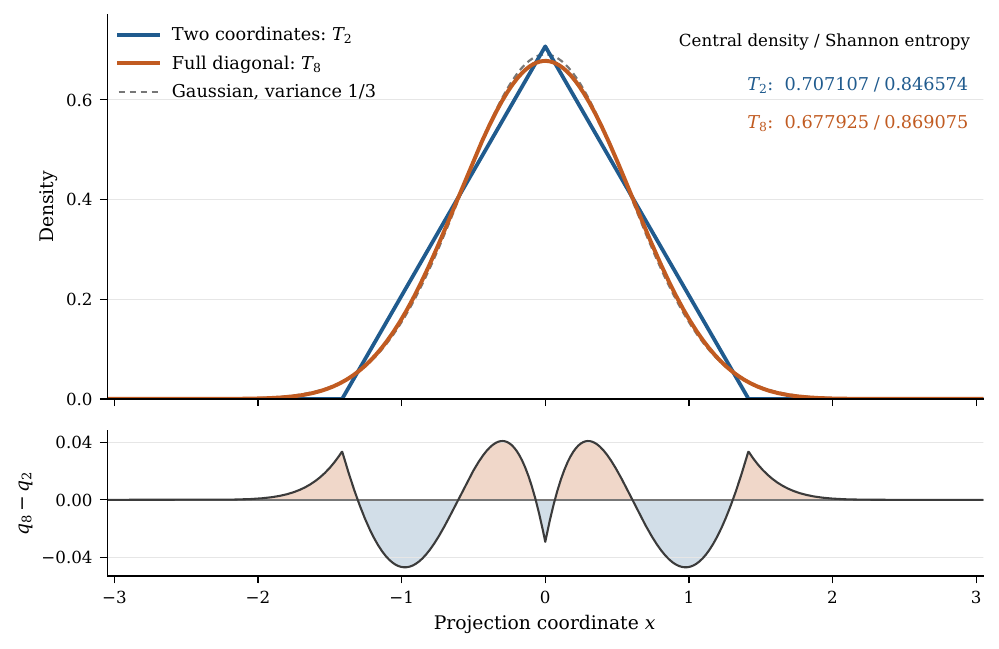}
\caption{Two projections of the eight-dimensional cube, normalized to variance
$1/3$. The two-coordinate direction gives the triangular density $q_2$,
whereas the full diagonal gives $q_8$. The former has the larger central
density, $q_2(0)=1/\sqrt2>151\sqrt2/315=q_8(0)$, and therefore the smaller
min-entropy $h_\infty=-\log\|q\|_\infty$. The diagonal has the larger
Shannon entropy: $h(T_2)=(1+\log2)/2\approx0.846574$ and
$h(T_8)\approx0.869075$ nats. The lower panel shows the redistribution of
density. The dashed curve is the Gaussian density with the same variance.
The numerical values illustrate the comparison; the proof does not use
the plot or its numerical quadrature.}
\label{fig:projection-density}
\end{figure}

\paragraph{Sections of the cube.}
For a unit direction, the coarea formula identifies the projection
density with the volumes of parallel hyperplane sections:
\[
 p_{S_a}(t)=2^{-n}\operatorname{vol}_{n-1}
 \bigl([-1,1]^n\cap\{x:\langle a,x\rangle=t\}\bigr).
\]
These densities are even and unimodal. For $n\ge2$, Ball's cube-slicing theorem
\cite{Ball1986} gives $p_{S_a}(0)\le1/\sqrt2$, with a maximizer
$a=(1,1,0,\ldots,0)/\sqrt2$. This maximizes the central section volume
and the density peak. Equivalently, it minimizes the order-infinity
R\'enyi entropy $h_\infty(S_a)=-\log\|p_{S_a}\|_\infty$.
Shannon entropy depends on the complete density rather than its peak.
For $n>2$, its maximizing direction uses all $n$ coordinates. The
comparison therefore separates a largest central section from the
entropy of the full family of parallel sections. Figure~\ref{fig:projection-density}
illustrates these two criteria at the same variance.

\paragraph{Earlier comparisons and Pang's theorem.}
Entropy maximization under coefficient balancing is known for other
classes of summands. Eskenazis, Nayar and Tkocz \cite[Theorem~8]{ENT2018}
proved the comparison for independent copies of a centered Gaussian
scale mixture, under the stated entropy finiteness assumptions.
Bartczak, Nayar and Zwara \cite[Theorem~1]{BNZ2021} proved the analogous
result for independent $\operatorname{Gamma}(\alpha)$ variables with
nonnegative coefficients when $n\alpha\ge1$.

For uniform summands, Lata\l a and Oleszkiewicz \cite{LO1995} developed
pairwise averaging of squared coefficients for moment comparisons.
Bartczak, Nayar and Zwara \cite[Section~8.3.1]{BNZ2021} observed that the
entropy comparison follows from a suitable sign of the third
logarithmic derivative of the diagonal density $q_n$. Their numerical
observations suggested this sign for $n\ge7$.
Pang's work \cite{Pang} establishes
\[
 (-\log q_n)'''(x)>0\qquad(0<x<\sqrt n)
\]
for every $n\ge150$, and proves the entropy comparison with equality
only at directions having equal coordinate magnitudes. His proof
combines exponential tilting with real Fourier inversion and estimates
uniform in the tilt. 

The oscillation of the uniform characteristic function is one reason
that general coefficient-comparison criteria do not immediately settle
the problem. For example, the principal Fourier criterion of Malliaris
\cite[Theorem~1.1]{Malliaris2025} assumes a positive, integrable Fourier
transform of the summand density, together with hypotheses on the test
function. The uniform transform $\sin t/t$ changes sign and does not
satisfy those assumptions. 

\subsection{Quantitative results and proof strategy}
To measure departure from the diagonal, define
\[
 \delta(a)=\sum_{i=1}^n\left(a_i^2-\frac1n\right)^2
          =\sum_{i=1}^na_i^4-\frac1n.
\]
This quantity vanishes exactly at the entropy-maximizing directions.
Let $q_n$ denote the density of $T_n$, and write $\KL(p\Vert q)$ for
relative entropy. The proof gives the following stability estimate.

\begin{theorem}\label{thm:stability}
For $\|a\|_2=1$,
\[
 h(T_n)-h(S_a)\ge
 \begin{cases}
 \delta(a)/1000,&3\le n\le6,\\[2pt]
 \displaystyle\frac{7\delta(a)}{180n}+\KL(p_a\Vert q_n),&n\ge7.
 \end{cases}
\]
Moreover, for every $\eta>0$ there is $N_\eta$ such that
\begin{equation}\label{eq:asymptotic-stability}
 h(T_n)-h(S_a)\ge
 \left(\frac3{50}-\eta\right)\frac{\delta(a)}n
 +\KL(p_a\Vert q_n),\qquad n\ge N_\eta.
\end{equation}
\end{theorem}

The asymptotic coefficient cannot be improved. Define
\[
 \mathsf C_n=\inf_{\substack{a\in\R^n,\ \|a\|_2=1\\\delta(a)>0}}
        \frac{n[h(T_n)-h(S_a)]}{\delta(a)}.
\]
\begin{theorem}\label{thm:sharp}
The optimal coefficients satisfy $\displaystyle\lim_{n\to\infty}\mathsf C_n=3/50$.
\end{theorem}

The proof has two finite components. For $7\le n\le149$, we verify a
quantitative version of the third-logarithmic-derivative criterion used
by Pang. The diagonal density is a polynomial on each spline interval,
so the required sign becomes a polynomial inequality with integer
coefficients. Its Bernstein coefficients give a finite exact certificate
on the whole interval. For $3\le n\le6$, we instead use one even
polynomial of degree eight as an entropy dual. Its expectation is
controlled globally by moments, while a local Taylor argument covers
the neighborhood where the dual approximation has a nonzero error.
Dimension two has an explicit formula.

To obtain Theorem~\ref{thm:stability} in every dimension, we extend the
finite certificate through $255$ and extract a quantitative margin from
Pang's Fourier estimates for $n\ge256$. The sharpness statement then
follows from the entropic Edgeworth expansion of
Bobkov--Chistyakov--G\"otze \cite{BCG}. Thus the completion of the
conjecture and the additional quantitative argument have separate
dependencies: the former needs Pang's theorem for $n\ge150$; the latter
uses his analytic estimates in a more detailed form.

\paragraph{Relation to the Courtade--Kumar argument.}
The local and global estimates in dimensions three through six follow
an organization also used in the author's work with Tran on the
Courtade--Kumar problem \cite[Sections~1.1 and~8]{KyTran2026}: use an
explicit entropy estimate near the extremizer and a different bound on
the complementary region, with enough remainder control for the two
regions to overlap. The precise starting point here is
\[
 h(q)-h(p)=\int_{\mathbb R}(p-q)\log q\,dx+\KL(p\Vert q),
\]
where the final term is the Bregman remainder of the convex functional
$p\mapsto\int p\log p$. In Section~\ref{sec:small}, polynomial moments
control the global comparison, and conditional variances control the
local remainder. This connection concerns the organization of entropy
deficit estimates; the present proof has no dependence on the
Courtade--Kumar theorem or its Boolean entropy-production inequalities.

\section{From logarithmic curvature to entropy}\label{sec:curvature-comparison}
For the middle range of dimensions, it is more efficient to compare
expectations under the diagonal density than to estimate the entropy of
each projection separately. Pang's averaging lemma supplies the required
comparison. We state the input explicitly, including its support condition.

\begin{lemma}[Pang's two-coefficient comparison {\cite[Lemma~2.1]{Pang}}]
\label{lem:pang-pair}
Let $W$ be even and $C^3$ on $(-R,R)$, with $W'''(x)\ge0$ for
$0<x<R$. Let $Y$ be independent of $U_1,U_2$, and suppose that $Y=0$
almost surely or that $Y$ has an even unimodal density. If $a\ge b\ge0$,
$c=\sqrt{(a^2+b^2)/2}$, $|Y|\le L$ almost surely, and $2c+L<R$, then
\[
 \E W(aU_1+bU_2+Y)\le\E W(cU_1+cU_2+Y).
\]
\end{lemma}

The lemma applies to any residual sum of independent centered uniform
variables, including sums with zero coefficients. Indeed, convolution
with a uniform density preserves even unimodality: if $g$ is even and
nonincreasing on $[0,\infty)$, then the derivative of
$(2b)^{-1}\int_{x-b}^{x+b}g(t)\,dt$ is nonpositive for almost every $x>0$.
Pang proves the comparison by applying a convex-order argument to the
squares of the two corresponding uniform sums. We use that lemma as an
input; the quantitative term below comes from subtracting a quartic.

\begin{proposition}[Quantitative curvature criterion]\label{prop:curvature-stability}
Let $n\ge5$ and $V_n=-\log q_n$. If, for some $b>0$,
\begin{equation}\label{eq:curvature-target}
 V_n'''(x)\ge\frac b n x,\qquad 0<x<\sqrt n,
\end{equation}
then every $a\in\mathbb R^n$ with $\|a\|_2=1$ satisfies
\begin{equation}\label{eq:curvature-entropy}
 h(T_n)-h(S_a)\ge\frac{b}{180n}\delta(a)+\KL(p_a\Vert q_n).
\end{equation}
Here $p_a$ denotes the density of $S_a$. In particular the entropy
inequality is strict unless $|a_i|=n^{-1/2}$ for every $i$.
\end{proposition}
\begin{proof}
Signs of the coefficients do not affect the distribution, so assume
$a_i\ge0$. Put
\[
 W(x)=V_n(x)-\frac{b}{24n}x^4.
\]
The spline density $q_n$ is positive and $C^{n-2}$ on its open support.
Thus $W$ is even and $C^3$, and \eqref{eq:curvature-target} gives
$W'''\ge0$ on the positive half of the support. Also $q_n$ is even
unimodal, so $q_n'(0)=0$ and $q_n''(0)\le0$. Consequently
$W''(0)=V_n''(0)\ge0$. It follows that $W$ is nondecreasing as a
function of $|x|$.

We first avoid the logarithmic singularities at the endpoints. Fix
$0<\rho<1$ and apply Lemma~\ref{lem:pang-pair} to $W(\rho x)$.
After averaging any two squared coefficients, the sum of the new
coefficients is at most $\sqrt n<\sqrt n/\rho$, by Cauchy--Schwarz.
The strict support condition of the lemma therefore holds at every
averaging step. Repeatedly average a largest and a smallest squared
coefficient. If their difference is $d$, then
$\sum_i(a_i^2-1/n)^2$ decreases by $d^2/2$. Since that sum is at most
$nd^2$, it tends to zero under iteration. All coefficients consequently
converge to $n^{-1/2}$, and bounded convergence gives
\[
 \E W(\rho S_a)\le\E W(\rho T_n).
\]
The function $W(\rho x)-W(0)$ increases to $W(x)-W(0)$ as
$\rho\uparrow1$, so monotone convergence yields
\[
 \E W(S_a)\le\E W(T_n).
\]
The right-hand side is finite: near an endpoint $q_n(x)$ is a constant
times $(\sqrt n-|x|)^{n-1}$. If $a$ is not flat, then
$\sum_i a_i<\sqrt n$, so $S_a$ stays in a compact subinterval of the
support of $q_n$. This also proves finiteness for coefficient vectors
having zeros. The flat case is immediate.

Because the support of $p_a$ is contained in that of $q_n$, the
cross-entropy identity gives
\[
 h(T_n)-h(S_a)
 =\E V_n(T_n)-\E V_n(S_a)+\KL(p_a\Vert q_n).
\]
Finally,
\[
 \E S_a^4=\frac13-\frac2{15}\sum_i a_i^4,
 \qquad \E T_n^4-\E S_a^4=\frac2{15}\delta(a).
\]
Substituting $V_n=W+bx^4/(24n)$ proves
\eqref{eq:curvature-entropy}. Since $\delta(a)\ge0$, with equality
exactly at the flat vectors, the equality assertion follows.
\end{proof}

\section{An exact spline certificate in the finite range}\label{sec:finite-certificate}
We now verify the curvature hypothesis without numerical differentiation.
The density of a sum of uniform variables is polynomial between consecutive
knots. Clearing the positive density denominators reduces the desired
inequality to positivity of an explicit integer polynomial on each cell.

\begin{theorem}[Finite curvature bound]\label{thm:finite-curvature}
For every integer $7\le n\le255$,
\[
 V_n'''(x)\ge\frac7n x,\qquad 0<x<\sqrt n.
\]
The assertion is certified by the exact integer computation described
below and implemented in \texttt{ent\_verify\_curvature.py}.
\end{theorem}
\begin{proof}
Let $f_n$ be the density of the sum of $n$ independent uniform variables
on $[0,1]$. For $k\le t\le k+1$, write $s=t-k$ and
\[
 (n-1)!f_n(k+s)=P_{n,k}(s),\qquad
 P_{n,k}(s)=\sum_{j=0}^k(-1)^j\binom nj(k+s-j)^{n-1}.
\]
On the positive half of the support of $q_n$, use the reflected
coordinate $t=(n-\sqrt n\,x)/2$. Symmetry of $f_n$ gives
\[
 q_n(x)=\frac{\sqrt n}{2(n-1)!}P_{n,k}(s).
\]
For a polynomial $P$, define
\[
 N(P)=P^2P'''-3PP'P''+2(P')^3.
\]
Differentiation, including the sign of $dt/dx$, gives
\[
 V_n'''(x)=\frac{n^{3/2}}8\frac{N(P_{n,k})(s)}{P_{n,k}(s)^3}.
\]
Since $P_{n,k}$ is positive in the interior of the support, the required
curvature inequality is equivalent to
\begin{equation}\label{eq:integer-curvature}
 R_{n,k}(s):=n^3N(P_{n,k})(s)
       -56(n-2k-2s)P_{n,k}(s)^3\ge0.
\end{equation}
All coefficients of $R_{n,k}$ are integers. The cells to check are
$0\le s\le1$ for $0\le k<n/2$, except that the last cell has width
$1/2$ when $n$ is odd.

For clarity, the exact positivity test is as follows. If
$R(s)=\sum_{i=0}^Da_is^i$, set
\begin{equation}\label{eq:bernstein-numerators}
 B_j=\sum_{i=0}^j\binom{D-i}{j-i}a_i,
 \qquad 0\le j\le D.
\end{equation}
Then
\[
 R(s)=\sum_{j=0}^D B_j s^j(1-s)^{D-j}.
\]
Equivalently, $B_j/\binom Dj$ is the usual degree-$D$ Bernstein
coefficient. Thus $B_j\ge0$ for every $j$ proves positivity on the
entire unit interval. On the last half cell, apply the same test to
$2^D R(s/2)$; its coefficients remain integers.

The supplied program computes every $P_{n,k}$ by the recurrence
\[
 P_{n,0}(s)=s^{n-1},\qquad
 P_{n,k+1}(s)=P_{n,k}(s+1)
       +(-1)^{k+1}\binom n{k+1}s^{n-1}.
\]
It forms \eqref{eq:integer-curvature} and checks
\eqref{eq:bernstein-numerators} exactly. Every numerator is nonnegative
in all $16\,372$ cells for $7\le n\le255$. This includes $5688$ cells
for $7\le n\le150$ and $10\,684$ further cells for $151\le n\le255$.
The former range already covers all middle dimensions required to
complete the entropy conjecture using Pang's theorem. The latter range
is used only for the uniform quantitative estimate.

The computation proves \eqref{eq:integer-curvature} on closed cells,
not merely at sampled points. Since $q_n$ is positive and $C^{n-2}$
on its open support, $V_n'''$ is continuous across the knots. The
cellwise inequalities therefore give the asserted bound everywhere
in $0<x<\sqrt n$.
\end{proof}

The program uses arbitrary-precision integer polynomial arithmetic.
For $7\le n\le12$, it additionally compares the recurrence with the
direct truncated-power sum and the fast Bernstein transform with
the defining binomial sums. The result file records every dimension,
the number of cells and coefficients, and a SHA-256 digest of the
full sequence of Bernstein numerators. These checks specify a finite
proof certificate whose sign decisions do not depend on rounding.

\section{The remaining four dimensions}\label{sec:small}
The curvature criterion is not needed in dimensions three through six.
Instead, we approximate the negative logarithm of the diagonal density by
an even polynomial. Its moments give a global comparison away from the
diagonal, while a local expansion handles the approximation error.

\begin{proposition}\label{prop:small}
For $3\le n\le6$ and $\sum_i a_i^2=1$,
\[
 h(T_n)-h(S_a)\ge \frac{\delta(a)}{1000}.
\]
\end{proposition}

We may assume $a_i\ge0$. Put $X_a=\sqrt3S_a$ and $X_*=\sqrt3T_n$,
so both variables have variance one. For a polynomial $W$ with positive
leading coefficient, set
\[
 Z_W=\int_{\mathbb R}e^{-W(x)}\,dx,\qquad
 \sigma_W=\log Z_W+\E W(X_*)-h(X_*).
\]
The quantity $\sigma_W$ is the relative entropy of $X_*$ with respect
to the density $e^{-W}/Z_W$. Nonnegativity of relative entropy gives
\begin{equation}\label{eq:small-gibbs}
 h(T_n)-h(S_a)
 \ge \E W(X_*)-\E W(X_a)-\sigma_W.
\end{equation}
Thus a moment comparison with a strictly positive margin suffices once
the coefficient vector is far enough from the diagonal.

\subsection{An octic moment comparison}
For $w_i=a_i^2$, write $p_j=\sum_iw_i^j$. Expanding independent even
moments yields
\begin{align*}
 \E X_a^4&=3-\frac65p_2,\\
 \E X_a^6&=15-18p_2+\frac{48}{7}p_3,\\
 \E X_a^8&=105-252p_2+192p_3+\frac{252}{5}p_2^2-\frac{432}{5}p_4.
\end{align*}
Consequently, if $W(x)=\sum_{j=1}^4c_jx^{2j}$, then
\[
 \E W(X_a)=c_1+3c_2+15c_3+105c_4-Ap_2+Bp_3+Cp_2^2-Dp_4,
\]
where
\[
 A=\frac65c_2+18c_3+252c_4,\quad
 B=\frac{48}{7}c_3+192c_4,\quad
 C=\frac{252}{5}c_4,\quad D=\frac{432}{5}c_4.
\]

\begin{lemma}\label{lem:small-schur}
Suppose $0<D<2C$ and
\[
 2(A-\lambda)\ge\max\{4C,\,3B+4C-4D\}.
\]
Then $\E W(X_*)-\E W(X_a)\ge\lambda\delta(a)$.
\end{lemma}
\begin{proof}
It suffices to show that $H(w)=\E W(X_a)+\lambda p_2$ is
Schur-concave on the probability simplex. For $w_i\ne w_j$, put
$s=w_i+w_j$ and $t=w_iw_j$. Since
$p_2\le s^2-2t+(1-s)^2$,
\begin{align*}
 \frac{\partial_iH-\partial_jH}{w_i-w_j}
 &=-2(A-\lambda)+3Bs+4Cp_2-4D(s^2-t)\\
 &\le-2(A-\lambda)+4C+(3B-8C)s+(8C-4D)s^2+(4D-8C)t\\
 &\le-2(A-\lambda)+\max\{4C,3B+4C-4D\}\le0.
\end{align*}
In the last line, the coefficient of $t$ is negative and the quadratic
in $s\in[0,1]$ is convex. Pairwise averaging now proves the claim;
continuity includes equal or zero coordinates.
\end{proof}

We use the following four polynomials. All terminating decimals in the
tables denote exact rational numbers.
\[
\begin{array}{c|rrrr}
 n&c_1&c_2&c_3&c_4\\\hline
 3&.345&.081&-.0226&.00221\\
 4&.4065&.0301&-.00563&.000518\\
 5&.43346&.01535&-.001769&.00017206\\
 6&.44749&.01001&-.0006493&.00007284
\end{array}
\]
The comparison slopes and upper bounds on the corresponding
$\sigma_n=\sigma_{W_n}$ are
\begin{equation}\label{eq:small-parameters}
\begin{array}{c|ccc}
 n&\lambda_n&\sigma_n\text{ upper bound }s_n&\delta_{0,n}\\\hline
 3&1/100&9/100000&1/100\\
 4&11/1000&1/50000&1/500\\
 5&1/100&1/200000&7/12500\\
 6&19/2000&7/5000000&1/6000
\end{array}
\end{equation}
The inequalities $\sigma_n<s_n$ are established by the interval
integrations specified in Appendix~\ref{app:small-integrals}.
For $n=4,5,6$, Lemma~\ref{lem:small-schur} applies: the smaller of
its two rational margins is, respectively,
\[
 \frac{589}{875000},\qquad
 \frac{10583}{5468750},\qquad
 \frac{21937}{87500000}.
\]

For $n=3$ the required moment comparison follows from a short polynomial
identity. Put $S=w_1+w_2+w_3$ and
\[
 M_k(w)=3^k(2k)!\sum_{\alpha_1+\alpha_2+\alpha_3=k}
       \prod_{i=1}^3\frac{w_i^{\alpha_i}}{(2\alpha_i+1)!},
 \qquad F(w)=\sum_{k=1}^4c_kM_k(w)S^{4-k}.
\]
Then $F=\E W_3(X_a)$ on $S=1$. Let $f_*=F(1/3,1/3,1/3)$ and
\[
 G=f_*S^4-F-\frac1{100}(p_2-S^2/3)S^2.
\]
After ordering $w_1\ge w_2\ge w_3$ and writing
$(w_1,w_2,w_3)=(x+y+z,y+z,z)$, direct expansion gives
\begin{align*}
 1312500G={}&4799x^4+58297x^3y+131163x^2y^2
                  +145732xy^3+72866y^4\\
 &+z(97398x^3+311562x^2y+350298xy^2+233532y^3)\\
 &+165465z^2(x^2+xy+y^2)\ge0.
\end{align*}
This proves the comparison with slope $\lambda_3=1/100$.
For every row of \eqref{eq:small-parameters},
$s_n\le(\lambda_n-1/1000)\delta_{0,n}$. Equation~\eqref{eq:small-gibbs}
therefore proves Proposition~\ref{prop:small} when $\delta\ge\delta_{0,n}$.

\subsection{A local comparison with fixed support}
We next control the error near the diagonal. The following estimate is
useful because it averages the singular third derivative against the
diagonal density rather than replacing it by its largest value.

Write $u=n^{-1/2}(1,\ldots,1)$ and parametrize a neighborhood of $u$ by
\[
 a=\frac{u+b}{\sqrt{1+z}},\qquad b\perp u,
 \quad z=\|b\|_2^2,\qquad T=u\cdot U,\quad Y=b\cdot U.
\]
For any $\gamma\ge\sqrt n\|b\|_\infty$ with $\gamma<1$, all coefficients of $T+Y$ are
positive and sum to $\sqrt n$, so its support is the same as that of $T$.
With $V=-\log q_n$ and $d(x)=\sqrt n-|x|$, relative entropy and scaling
give
\begin{equation}\label{eq:small-fixed-support}
 h(S_a)-h(T_n)
 \le \E[V(T+Y)-V(T)]-\tfrac12\log(1+z).
\end{equation}

\begin{lemma}\label{lem:small-conditional}
For almost every $T$,
\[
 \E[Y\mid T]=0,\qquad |Y|\le\gamma d(T),\qquad
 \E[Y^2\mid T]\le\frac{z\,d(T)^2}{n+1}.
\]
Moreover $\sqrt n\|b\|_\infty\le\sqrt{n-1}\|b\|_2$, so
$\gamma=\sqrt{n-1}\|b\|_2$ is admissible whenever it is less than one.
\end{lemma}
\begin{proof}
Exchangeability gives the conditional mean. Set
$X_i=1-\operatorname{sgn}(T)U_i$, choosing either sign when $T=0$.
Conditionally on $T$, the vector $X$ is uniform on
\[
 \{x_i\ge0:\ \textstyle\sum_i x_i=s,\ x_i\le2\},
 \qquad s=\sqrt n\,d(T).
\]
Since $\sum_i b_i=0$, $|Y|=|\sum_i b_iX_i|\le\|b\|_\infty s$.

To bound its conditional variance, compare this capped simplex with the
uncapped simplex $\{x_i\ge0:\sum_i x_i=s\}$. The ratio of their
one-coordinate marginal densities is proportional to
\[
 \mathbf1_{\{x\le2\}}
 \Pr\left\{\max_{1\le j\le n-1}D_j\le\frac2{s-x}\right\},
 \qquad D\sim\operatorname{Dirichlet}(1,\ldots,1).
\]
This ratio is nondecreasing before the cap and zero after it. Both
marginals have mean $s/n$. Thus their density difference has sign pattern
$-,+,-$, with degenerate cases allowed, and the capped marginal is
dominated by the uncapped marginal in convex order. Indeed, subtract
from a convex function the affine line through the two crossing points:
the resulting function is nonnegative outside them and nonpositive
between them, giving the stated integral comparison. In particular,
\[
 \E X_1^2\le\frac{2s^2}{n(n+1)},\qquad
 \Var(X_1)\le\frac{(n-1)s^2}{n^2(n+1)}.
\]
Exchangeability and the deterministic sum imply that the covariance
matrix on the zero-sum subspace is scalar, with eigenvalue
$n\Var(X_1)/(n-1)\le s^2/[n(n+1)]$. This proves the variance bound.
Finally, $\sum_i b_i=0$ implies
$\|b\|_\infty^2\le(n-1)\|b\|_2^2/n$.
\end{proof}

Let $f_j$ be the density of the sum of $j$ independent uniforms on $[0,1]$,
extended by zero outside $[0,j]$. Define
\begin{equation}\label{eq:small-kappa}
 \kappa_n=1-n\int_0^n[f_{n-2}(t-1)-f_n(t)]\log f_n(t)\,dt.
\end{equation}
For $n\ge4$, this constant satisfies
\begin{equation}\label{eq:small-hessian}
 \E[Y^2V''(T)]=(1-\kappa_n)z.
\end{equation}
To see this, differentiate the characteristic function of $T+\theta Y$
twice at $\theta=0$. The first derivative vanishes and the second is
\[
 nz\bigl[\operatorname{sinc}(t/\sqrt n)^n
             -\operatorname{sinc}(t/\sqrt n)^{n-2}\bigr].
\]
Fourier inversion and integration against $V$ give
\eqref{eq:small-hessian}, with the constant normalization in $\log q_n$
canceling because the two densities have equal mass. The unbounded test
function causes no endpoint difficulty: for sufficiently small $\theta$,
the perturbed density near either fixed endpoint equals
$C_\theta d(x)^{n-1}$, with $C_\theta$ twice continuously differentiable.
This dominates the logarithmic singularity uniformly. On the remaining
compact interval, the truncated-power density formula is twice
continuously differentiable in $\theta$ for $n\ge4$.

For $n\ge5$ and $0\le\gamma<1$, put
\[
 M_n(\gamma)=2\int_0^{n/2}f_n(t)t^3
  \sup_{(1-\gamma)t\le v\le\min(n/2,(1+\gamma)t)}
                |(\log f_n)'''(v)|\,dt.
\]
For these dimensions $V_n$ is $C^3$ on its open support, so the third
derivative in this supremum is continuous across the knots.

\begin{lemma}\label{lem:small-local}
For $n\ge5$ and $\gamma\ge\sqrt{n-1}\sqrt z$ with $\gamma<1$,
\[
 h(S_a)-h(T_n)
 \le z\left[-\frac{\kappa_n}{2}+\frac z4
                    +\frac{\gamma M_n(\gamma)}{6(n+1)}\right].
\]
\end{lemma}
\begin{proof}
The case $z=0$ is immediate, so assume $z>0$. Define the function
\[
 H_\gamma(t)=\sup_{\substack{(1-\gamma)d(t)\le d(x)\\
              d(x)\le\min(\sqrt n,(1+\gamma)d(t))}}|V'''(x)|.
\]
Every point of the segment $T+sY$, $0\le s\le1$, belongs to this set,
because support distance is Lipschitz and $|Y|\le\gamma d(T)$.
The function $H_\gamma(T)$ depends only on $T$. Taylor's theorem and
Lemma~\ref{lem:small-conditional} therefore bound the expected remainder by
\[
 \frac16\E[|Y|^3H_\gamma(T)]
 =\frac16\E[\E(|Y|^3\mid T)H_\gamma(T)]
 \le\frac{\gamma z}{6(n+1)}\E[d(T)^3H_\gamma(T)]
 =\frac{\gamma z}{6(n+1)}M_n(\gamma).
\]
Here $\E(|Y|^3\mid T)\le\gamma z d(T)^3/(n+1)$, and the substitution
$t=\sqrt n\,d(T)/2$ cancels the derivative scaling in the last equality.
The linear term
vanishes by conditional mean zero. Combining the quadratic term
\eqref{eq:small-hessian} with \eqref{eq:small-fixed-support} and
$\log(1+z)\ge z-z^2/2$ proves the claim.
\end{proof}

For $n=5,6$, the required scalar bounds are
\[
\begin{array}{c|cccc}
 n&r_n&\gamma_n&M_n(\gamma_n)\text{ upper bound}&\kappa_n\text{ lower bound}\\\hline
 5&29/1000&29/500&77/10&27/1000\\
 6&7/400&1/25&13/2&17/1000
\end{array}
\]
Appendix~\ref{app:small-integrals} specifies their verification.
If $z\le r_n^2$, then
\[
 \delta(a)=\frac{4z/n+(4/\sqrt n)\sum_i b_i^3+\sum_i b_i^4-z^2/n}
                 {(1+z)^2}
 \le\left(\frac4n+2r_n+r_n^2\right)z.
\]
The last inequality uses
$\|b\|_\infty\le\sqrt{(n-1)/n}\sqrt z$ and
$4\sqrt{n-1}/n\le2$. After subtracting the target $\delta/1000$,
Lemma~\ref{lem:small-local} leaves the positive rational margins
\[
 \frac{228181}{9000000000}\quad(n=5),\qquad
 \frac{1714707}{1120000000}\quad(n=6).
\]

It remains to check that this local region reaches the thresholds in
\eqref{eq:small-parameters}. If $\delta\le1/(16n^2)$, then
$a_i^2\ge3/(4n)$. Since $(1+\sqrt3/2)^2>10/3$,
\[
 \|a-u\|_2^2
 =\sum_i\frac{(a_i^2-1/n)^2}{(a_i+1/\sqrt n)^2}
 \le\frac{3n\delta}{10},\qquad
 z=\langle a,u\rangle^{-2}-1
 \le\left(1-\frac{3n\delta}{20}\right)^{-2}-1.
\]
At $\delta=\delta_{0,n}$ the latter bound is
$2099559/2497900441<r_5^2$ for $n=5$ and
$119991/399880009<r_6^2$ for $n=6$.

For dimensions three and four, direct primitives give the convenient
local bounds
\[
 h(T_3)-h(S_a)\ge\delta/720\quad(\delta\le1/100),\qquad
 h(T_4)-h(S_a)\ge\delta/225\quad(\delta\le1/500).
\]
Their proofs are given in Appendices~\ref{app:small-three}
and~\ref{app:small-four}. These neighborhoods and the two preceding
local estimates overlap the global regions from
\eqref{eq:small-gibbs}, completing the proof of
Proposition~\ref{prop:small}.

\section{Completion of the entropy comparison}\label{sec:completion}
We now prove Theorem~\ref{thm:main}. Dimension one is immediate. In
dimension two, symmetry permits $a\ge b\ge0$ with $a^2+b^2=1$.
Put $r=b/a$. Direct integration of the trapezoidal density gives
\[
 h(aU_1+bU_2)=\log2-\frac12\log(1+r^2)+\frac r2.
\]
Consequently
\[
 h(T_2)-h(aU_1+bU_2)
 =\int_r^1\frac{(1-s)^2}{2(1+s^2)}\,ds,
\]
which is positive unless $r=1$. For $3\le n\le6$, the octic comparison
and local estimates of the preceding section give a gap of at least
$\delta(a)/1000$. For $7\le n\le149$, the finite curvature certificate
and the comparison principle in Section~\ref{sec:curvature-comparison} give a gap of
at least $7\delta(a)/(180n)$. Finally, Pang's theorem \cite[Theorem~1.1]{Pang}
gives the result and its equality case for $n\ge150$. In every range,
equality is equivalent to $\delta(a)=0$.

The exclusion of dimension two from the linear stability bound is
necessary. In the preceding parametrization,
\[
 \delta(a)=\frac{(1-r^2)^2}{2(1+r^2)^2},\qquad
 h(T_2)-h(S_a)\sim\frac{\sqrt2}{6}\delta(a)^{3/2}
 \quad(r\uparrow1).
\]

\section{Quantitative bounds in large dimensions}\label{sec:large}
Pang's theorem already supplies the entropy comparison for $n\ge150$.
We now retain the size of the remainder in his Fourier argument to obtain
a stability estimate. This refinement is separate from the proof of the
entropy maximum.

\begin{proposition}\label{prop:large-curvature}
For $n\ge256$ and $0<x<\sqrt n$,
\[
 V_n'''(x)>\frac7n x.
\]
Moreover, for every $\varepsilon\in(0,1/10]$, there is an integer
$N_\varepsilon$ such that, uniformly in $0<x<\sqrt n$,
\[
 \frac{nV_n'''(x)}x\ge\frac{54}{5}(1-\varepsilon)
 \qquad(n\ge N_\varepsilon).
\]
\end{proposition}

\subsection{Exponential tilting and a scalar inequality}
Tilting moves the point at which the density is differentiated to the
mean of a new distribution. Let
\[
 K(t)=\log\E e^{tU_1}=\log\frac{\sinh t}{t},\qquad
 m=K'(t),\quad v=K''(t),\qquad t>0,
\]
and let $U_t$ have density $e^{tu-K(t)}/2$ on $[-1,1]$.
Set $Y=(U_t-m)/\sqrt v$, let $W=n^{-1/2}\sum_{i=1}^nY_i$, and write
$w$ for its density and $\mu_3=\E Y^3=K'''(t)/v^{3/2}<0$.
Pang's density transformation \cite[Section~3]{Pang}, differentiated at a
fixed tilt, gives
\begin{equation}\label{eq:large-tilt}
 V_n'''(\sqrt n\,m)=v^{-3/2}(-\log w)'''(0).
\end{equation}
Since $K'$ increases from $0$ to $1$, every positive point of the support
is obtained in this way. We shall use the following sharp estimate; its
proof is given in Appendix~\ref{app:langevin-paper}.

\begin{lemma}\label{lem:langevin-paper}
For every $t>0$,
\[
 \frac{-K'''(t)}{K'(t)K''(t)^3}>\frac{54}{5},
 \qquad
 \lim_{t\downarrow0}\frac{-K'''(t)}{K'(t)K''(t)^3}=\frac{54}{5}.
\]
\end{lemma}

\subsection{The Fourier estimates retained from Pang}
We state the inputs explicitly to distinguish them from the constants
extracted below. Put
\[
 r_n(u)=\left(1+\frac{u^2}{n}\right)^{-(n-1)/2},\quad
 C_n(u)=\E\cos(uW),\quad
 S_n(u)=\frac{\sqrt n}{|\mu_3|}\E\sin(uW).
\]
Pang proves $|\mu_3|\le2$ and the following bounds for all real $u$
\cite[Lemma~4.1, Corollary~4.4, (45)--(46)]{Pang}:
\[
 |C_n(u)|\le r_n(u),\qquad |S_n(u)|\le\frac{|u|^3}{6}r_n(u).
\]
For $0\le u\le\sqrt n/2$ his Lemma~4.6 gives
\begin{align}
 C_n(u)&\ge e^{-u^2/2}\left(1-\frac{u^4}{7n}\right)
                   -\frac{u^6}{12n}r_n(u),\label{eq:large-cos}\\
 S_n(u)&\ge\frac{u^3}{6}e^{-u^2/2}
       \left(1-\frac{11u^2}{10n}-\frac{u^4}{7n}\right)
                   -\frac{u^9}{216n}r_n(u).\label{eq:large-sin}
\end{align}
All applications here have $n\ge256$, within the stated range of these
lemmas. Fourier inversion gives
\begin{align*}
 w(0)&=\frac1\pi\int_0^\infty C_n(u)\,du,&
 w''(0)&=-\frac1\pi\int_0^\infty u^2C_n(u)\,du,\\
 w'(0)&=\frac{|\mu_3|}{\pi\sqrt n}\int_0^\infty uS_n(u)\,du,&
 w'''(0)&=-\frac{|\mu_3|}{\pi\sqrt n}\int_0^\infty u^3S_n(u)\,du.
\end{align*}

To integrate the bounds without dimension-independent rounding, define
\[
 M_{2j}(n)=\sqrt{\frac2\pi}\int_0^\infty u^{2j}r_n(u)\,du.
\]
Pang's Appendix~B gives the moment recurrence and the elementary bound
\[
 M_0(n)\le\mathfrak m_0:=\frac{2n-3}{2(n-3)},\qquad
 M_{2j}(n)\le\mathfrak m_{2j}:=
 \mathfrak m_0\prod_{\ell=1}^j\frac{(2\ell-1)n}{n-2\ell-2}
 \quad(1\le j\le6).
\]
For later use set
\[
 T_0=\frac{64(15+\mathfrak m_6)}{n^3},\quad
 T_4=\frac{256(10395+\mathfrak m_{12})}{n^4},\quad
 T_6=\frac{64(10395+\mathfrak m_{12})}{n^3}.
\]
These bound, respectively, the integrals of
$e^{-u^2/2}+r_n(u)$, $u^4(e^{-u^2/2}+r_n(u))$, and
$u^6(e^{-u^2/2}+r_n(u))$ over $u>\sqrt n/2$, with normalization
$\sqrt{2/\pi}$. For example on this tail $1\le64u^6/n^3$ and
$u^4\le256u^{12}/n^4$.

Integrating \eqref{eq:large-cos} and \eqref{eq:large-sin} now yields
\[
 \sqrt{2\pi}\,w(0)\ge A,
 \qquad \frac{\sqrt{2\pi n}\,w'(0)}{|\mu_3|}\ge B,
\]
where
\[
 A=1-\frac3{7n}-\frac{\mathfrak m_6}{12n}-T_0,
 \qquad
 B=\frac12-\frac{21}{4n}-\frac{\mathfrak m_{10}}{216n}-\frac{T_4}{6}.
\]
The absolute bounds also give
\[
 \sqrt{2\pi}\,w(0)\le\mathfrak m_0,\quad
 \sqrt{2\pi}\,|w''(0)|\le\mathfrak m_2,\quad
 \frac{\sqrt{2\pi n}\,|w'(0)|}{|\mu_3|}\le\frac{\mathfrak m_4}{6}.
\]
Thus, when $A,B>0$, $w'(0)>0$ and
\[
 -\frac{3w''(0)}{w(0)}+\frac{2w'(0)^2}{w(0)^2}
 \le\mathcal C(n):=
 \frac{3\mathfrak m_2}{A}+\frac8n\left(\frac{\mathfrak m_4}{6A}\right)^2.
\]

\subsection{A quantitative signed integral}
Fix $3<c\le31/10$, with $n>4c$, and write
\[
 J_c=\sqrt{\frac2\pi}\int_0^\infty u(u^2-c)S_n(u)\,du
 =-\frac{\sqrt{2\pi n}}{|\mu_3|}\,[w'''(0)+cw'(0)].
\]
Replacing $S_n(u)$ by $u^3e^{-u^2/2}/6$ gives the main term $(5-c)/2$.
We bound the error on the three intervals separated by $\sqrt c$ and
$\sqrt n/2$, following the sign split in Pang's Proposition~4.8.

On $[0,\sqrt c]$, the weight is nonpositive. The absolute sine bound and
$\log(1+s)\ge s-s^2/2$ give
\[
 S_n(u)\le\frac{u^3}{6}r_n(u)
 \le e^{x_n}\frac{u^3}{6}e^{-u^2/2},
 \qquad x_n=\frac{c/2+c^2/4}{n}.
\]
The resulting loss is at most $I_c(e^{x_n}-1)$, where
\[
 I_c=\sqrt{\frac2\pi}\int_0^{\sqrt c}
          \frac{u^4(c-u^2)}6e^{-u^2/2}\,du
 \le\sqrt{\frac2\pi}\frac{c^{7/2}}{105}<\frac25.
\]
On $[\sqrt c,\sqrt n/2]$ use \eqref{eq:large-sin}. Extending its
nonnegative error integrand to $[\sqrt c,\infty)$ bounds the loss by
$H_c/n$, with
\[
 H_c=\frac{167-21c}{4}
 +\frac{\mathfrak m_{12}}{216}\left(1-\frac{c(n-14)}{11n}\right)
 +\frac{11}{10}.
\]
Indeed the integrals over the full half-line give the first two terms:
the Gaussian moments give $(167-21c)/4$, and
$M_{12}-cM_{10}=M_{12}[1-c(n-14)/(11n)]$.
The correction for the discarded interval $[0,\sqrt c]$ is at most
\[
 R_c\le\sqrt{\frac{2c}\pi}
 \sum_{(j,b)\in\mathcal J}
 \frac{2b\,c^{(j+2)/2}}{(j+1)(j+3)}<\frac{11}{10},
 \quad\mathcal J=\{(6,11/60),(8,1/42),(10,1/216)\}.
\]
The last inequality and the bound for $I_c$ hold at $c=31/10$ by squaring
and using $\pi>157/50$; both right-hand bounds increase with $c$.
On the final tail the loss is at most $T_6/6$, since
$0<u(u^2-c)\le u^3$ and $|S_n(u)|\le u^3r_n(u)/6$.

For $0<x<1$, the exponential series gives
\[
 e^x-1\le\mathcal R(x):=x+\frac{x^2}{2}+\frac{x^3}{6}+\frac{x^4}{24}
                  +\frac{x^5/120}{1-x/6}.
\]
We have therefore proved
\[
 J_c\ge F_c(n):=\frac{5-c}{2}-\frac25\mathcal R(x_n)
                          -\frac{H_c}{n}-\frac{T_6}{6}.
\]
Suppose that $A,B,F_c>0$ and $\mathcal C(n)<c$. The identity
\[
 (\log w)'''(0)=\frac{w'''(0)}{w(0)}+
 \frac{w'(0)}{w(0)}
 \left[-\frac{3w''(0)}{w(0)}+\frac{2w'(0)^2}{w(0)^2}\right]
\]
then implies, using $w'(0)>0$ and the upper bound on $w(0)$,
\[
 (-\log w)'''(0)\ge\frac{|\mu_3|}{\sqrt n}
                          \frac{F_c(n)}{\mathfrak m_0}.
\]
Combining this with \eqref{eq:large-tilt} and
Lemma~\ref{lem:langevin-paper} gives
\begin{equation}\label{eq:large-final}
 \frac{nV_n'''(x)}x\ge\frac{54}{5}\frac{F_c(n)}{\mathfrak m_0},
 \qquad 0<x<\sqrt n.
\end{equation}

At $n=256$, $c=31/10$, direct rational arithmetic proves $A,B,F_c>0$,
$\mathcal C<c$, and
\[
 \frac{54F_c}{5\mathfrak m_0}
 =\frac{23157436997168015517032512478271549}
        {3302784924207803898265600000000000}>7.
\]
The verifier \texttt{ent\_verify\_large.py} records these rational checks.
All $\mathfrak m_{2j}$ and $T_j$ decrease with $n$; $A,B$ increase,
and $\mathcal C$ decreases. The positive factor
$1-c/11+14c/(11n)$ also decreases. Hence every loss in $F_c$ decreases,
and $F_c/\mathfrak m_0$ increases once $F_c>0$. This proves the first
part of Proposition~\ref{prop:large-curvature} for every $n\ge256$.
Finally fix $c=3+\varepsilon$ with $0<\varepsilon\le1/10$.
Then $\mathcal C(n)\to3$, $\mathfrak m_0\to1$, and
$F_c(n)\to1-\varepsilon/2$. Formula~\eqref{eq:large-final} therefore
proves the second part of the proposition.

Combining Proposition~\ref{prop:small}, Theorem~\ref{thm:finite-curvature},
Proposition~\ref{prop:large-curvature} and the curvature criterion proves
the first assertion of Theorem~\ref{thm:stability}. Given $\eta>0$, choose
$0<\varepsilon\le1/10$ with $3\varepsilon/50<\eta$. The second assertion
of Proposition~\ref{prop:large-curvature}, followed by
Proposition~\ref{prop:curvature-stability}, proves
\eqref{eq:asymptotic-stability}.

\section{The sharp asymptotic stability coefficient}\label{sec:sharp}
Proposition~\ref{prop:large-curvature} and the curvature comparison in
Proposition~\ref{prop:curvature-stability} give
$\liminf_{n\to\infty}\mathsf C_n\ge3/50$.
We next obtain the matching upper bound from a single explicit sequence
of coefficients.

\begin{proof}[Proof of Theorem~\ref{thm:sharp}]
Let $\gamma$ be the standard Gaussian law and
$D_n=D(\sqrt3T_n\Vert\gamma)$. The entropic Edgeworth theorem of
Bobkov, Chistyakov and G\"otze \cite[Theorem~1.1]{BCG} applies to
$X_i=\sqrt3U_i$: all absolute moments are finite and
$D(X_1\Vert\gamma)<\infty$. Their third cumulant vanishes and their
fourth cumulant is $\kappa_4=9/5-3=-6/5$.
Taking the moment parameter $s=8$ in that theorem yields
\[
 D_n=\frac{\kappa_4^2}{48n^2}+\frac{c_3}{n^3}+o(n^{-3})
     =\frac{3}{100n^2}+\frac{c_3}{n^3}+o(n^{-3})
\]
for a fixed constant $c_3$. Only its existence is needed.
For $a^{(n)}=((n-1)^{-1/2},\ldots,(n-1)^{-1/2},0)$,
$S_{a^{(n)}}$ has law $T_{n-1}$ and
$\delta(a^{(n)})=1/[n(n-1)]$. Therefore
\[
 \mathsf C_n\le n^2(n-1)[h(T_n)-h(T_{n-1})]
               =n^2(n-1)(D_{n-1}-D_n)\longrightarrow\frac3{50}.
\]
The remainder is sufficient: if $r_n=o(n^{-3})$, then
$n^3(r_{n-1}-r_n)=o(1)$ by the triangle inequality. Together with the
lower bound this proves the claim.
\end{proof}
The theorem determines the limiting coefficient, but does not identify
an optimizer or the optimal coefficient at any fixed dimension.

\section{Reproducibility}\label{sec:reproduce}
The finite verification uses two kinds of rigorous arithmetic. Polynomial
identities, coefficient signs, overlap inequalities and the
large-dimensional scalar closure use integers or rational numbers.
The remaining one-dimensional integrals are enclosed by Arb ball arithmetic \cite{Johansson2017arb} through \texttt{python-flint}. Their verification includes
quadrature remainder and tail bounds; numerical sampling is not used to
infer a global inequality. Appendix~\ref{app:small-integrals} specifies these
bounds.

The \href{https://github.com/vukhacky/ent-conjecture/}{GitHub repository}
contains the verification scripts. The accompanying archive includes
these scripts, their outputs, the manuscript and its LaTeX sources,
and the figure-generation script. From the extracted archive directory,
run \texttt{python3 code/run\_all.py} to verify the certificates and write
the results to \texttt{results/}. Exact dependency versions and compilation
instructions are provided in \texttt{README.md}.
The density figure is an illustration computed by ordinary numerical
quadrature; it is not used in any proof.

\paragraph{Use of AI tools.}
The author used ChatGPT for language editing, brainstorming, and coding
assistance. The author takes responsibility for the mathematical claims,
the proofs, and the accompanying computations.

\appendix
\section{A local comparison in dimension three}\label{app:small-three}

Write $r=1/\sqrt3$, $R=\sqrt3$, and let $q$ be the density of $T_3$:
\[
q(x)=\begin{cases}
(3\sqrt3/8)(1-x^2),&0\le x\le r,\\
(3\sqrt3/16)(R-x)^2,&r\le x<R,
\end{cases}
\qquad q(-x)=q(x).
\]
Set $V=-\log q$, $h=h(T_3)$,
$m=\frac13\sum_i a_i$, $z_i=a_i-m$, $u=\rho^2=\sum_i z_i^2$.
Then $m=r+d=r\sqrt{1-u}$. For $\rho\le1/10$,
\[
d\le-ru/2,\quad |d|\le3u/10,\quad |z_i|\le\sqrt{2/3}\rho,
\quad p=z_1z_2z_3,\quad |p|<\rho^3/7.
\]
Also $\sum z_i^3=3p$ and $\sum z_i^4=u^2/2$.

Let $F(x)=\frac12\int_0^x(x-t)^2V(t)\,dt$ for $x\ge0$ and extend
$F$ oddly. Three integrations of the product-uniform density give, with
$A=3m$,
\[
\E V(S_a)=\frac{F(A)-\sum_iF(A-2a_i)}{4a_1a_2a_3}.
\]
The arguments $A-2a_i=r+d-2z_i$ lie in $(0,2r)$, where
$0\le V''\le6$ almost everywhere; $V'$ is continuous at $r$.
The total fourth-order Taylor error at $r$ is consequently bounded by
\[
E_5\le\frac1{20}\max_i|d-2z_i|\sum_i(d-2z_i)^4.
\]
At the outer endpoint $F,F',F''$ are finite and $V>0$, whence the
one-sided integral remainder gives
$F(A)\le F(R)+3dF'(R)+\frac92d^2F''(R)$.

The constants needed below are
\[
\begin{gathered}
v=\log(4/\sqrt3),\qquad
h=\frac76+2\log2-\frac12\log3-\sqrt3\operatorname{arctanh}r,\\
J=rv+2r-2\operatorname{arctanh}r,
\qquad F''(R)=J+2r(v+2),\\
B=-2rh+2J,\quad C=12rh+\frac32J-\frac92F''(R),
\quad e=h-v,\quad K=4h+v/2.
\end{gathered}
\]
The identity
$12hr^2+3(F'(r)-F'(R))=-4/3$ follows, for example, by differentiating
along the radial path $a_i=r+d$ and using $\E[T_3V'(T_3)]=1$,
an integration-by-parts identity. Together with the bounds in
Appendix~\ref{app:small-integrals},
\[
B+2r/3>3/100,\quad -10<C<0,\quad 0<e<27/1000,\quad0<K<4,
\]
it controls all coefficients. Multiplying by $4a_1a_2a_3$ and using
$a_1a_2a_3=(r+d)^3-(r+d)u/2+p$ yields
\begin{align*}
D_3:=4a_1a_2a_3[h-\E V(S_a)]\ge{}&-\frac43d+Bu+Cd^2-2edu+4ep+Kd^3\\
&+\sqrt3\left(\frac{u^2}{3}+d^2u-4dp+\frac{d^4}{8}\right)-E_5.
\end{align*}
For $\rho\le1/10$ the coefficient $4e-4\sqrt3d$ of $p$ lies in
$(0,13/100)$. Moreover,
\[
\max_i|d-2z_i|\le\frac{17}{10}\rho,
\quad \sum_i(d-2z_i)^4=3d^4+24d^2u-96dp+8u^2\le9\rho^4.
\]
After discarding only nonnegative terms,
\[
\frac{D_3}{u}\ge\frac3{100}-\frac9{10}\rho^2-\frac{27}{250}\rho^4
                  -\frac{13}{700}\rho-\frac{153}{200}\rho^3>\frac1{60}.
\]
The right-hand side decreases with $\rho$; at $\rho=1/10$ its margin
over $1/60$ is $178541/105000000>0$.

If $\delta\le1/100$, then $a_i^2\ge7/30>(12/25)^2$ and
$r>13/25$, so $a_i+r>1$ and
\[
\rho^2\le\sum_i(a_i-r)^2
 =\sum_i\frac{(a_i^2-1/3)^2}{(a_i+r)^2}\le\delta\le1/100.
\]
Also $\delta\le\sum_i(a_i^2-m^2)^2\le3\rho^2$, because
$a_i\le1$, $m\le1/\sqrt3$ and $(1+1/\sqrt3)^2<3$.
The support of $S_a$ is contained in $[-R,R]$, so
$h(S_a)=\E V(S_a)-\KL(p_a\Vert q)\le\E V(S_a)$.
Using the coarse bound $a_1a_2a_3\le1$, we obtain
\[
h(T_3)-h(S_a)\ge\frac{\rho^2}{240}\ge\frac{\delta}{720}.
\]
This proves the local comparison in dimension three. The case $\rho=0$
is the diagonal and follows directly.

\section{A local comparison in dimension four}\label{app:small-four}

The even density of $T_4$ is
\[
q(x)=\begin{cases}2/3-x^2+x^3/2,&0\le x\le1,\\
(2-x)^3/6,&1\le x<2.
\end{cases}
\]
Write $V=-\log q$, $h=h(T_4)$, $m=\frac14\sum_i a_i=1/2+d$,
$z_i=a_i-m$, $u=\rho^2=\sum_i z_i^2$, and $s_j=\sum_i z_i^j$.
For $\rho\le1/20$,
\[
d\le-u/4,\quad |d|\le13u/50,\quad |s_3|\le\rho^3,\quad s_4\le u^2,
\quad \prod_i a_i=m^4-m^2u/2+ms_3/3+u^2/8-s_4/4.
\]
Let $F(x)=\frac16\int_0^{|x|}(|x|-t)^3V(t)\,dt$, $A=4m$, and
$t_j=A-2a_1-2a_j=-2(z_1+z_j)$ for $j=2,3,4$.
Four integrations give
\[
\E V(S_a)=\frac{F(A)-\sum_{i=1}^4F(A-2a_i)+\sum_{j=2}^4F(t_j)}
                     {8\prod_i a_i}.
\]
All inner arguments $A-2a_i=1+2d-2z_i$ lie in $[3/4,5/4]$,
where $|V'|\le4$. Taylor expansion at $1$ through degree four therefore
has total error
\[
E_5\le\frac1{30}\max_i|2d-2z_i|\sum_i(2d-2z_i)^4
       \le\frac65\rho^5.
\]
Indeed the maximum is at most $21\rho/10$ and the fourth-power sum
$64d^4+96d^2u-64ds_3+16s_4$ is at most $17\rho^4$.
The outer endpoint has the favorable bound
$F(A)\le F(2)+4dF'(2)+8d^2F''(2)$ since $F'''>0$.
Furthermore $|t_j|\le2\rho\le1/10$ and $V(1/10)<421/1000$, so
\[
\sum_{j=2}^4F(t_j)\le\frac{421}{24000}\sum_{j=2}^4t_j^4
\le\frac{421}{1000}u^2,
\qquad \sum_{j=2}^4t_j^4=24u^2-32s_4.
\]

Put
\[
v=\log6,\quad J_2=\int_0^1(1-x)V(x)\,dx,\quad
J_3=\int_0^1V(x)\,dx,\quad F''(2)=J_2+J_3+v/2+3/4.
\]
The interval bounds in Appendix~\ref{app:small-integrals} give
\[
\begin{gathered}
\alpha:=1/4-h+2J_2>11/1000,\quad -10<C:=12h+8J_2-8F''(2)<0,\\
0<e:=h-J_3<13/1000,\quad0<K:=16h+(16/3)J_3<19,\\
v>h>86/100,\quad (8/3)(v-h)<5/2,\quad -54/100<(2/3)v-2h<0.
\end{gathered}
\]
The coefficient of $d$ below is exactly $-1$. For example this follows
by differentiating from the left along $a_i=1/2+d$ and using
$\E[T_4V'(T_4)]=1$, which follows by integration by parts;
$xq(x)$ vanishes at both endpoints.
Expansion and cancellation of the constant term yield
\begin{align*}
D_4:=8\prod_i a_i[h-\E V(S_a)]\ge{}&-d+(-h+2J_2)u+Cd^2-4edu
                        +(4/3)e s_3+Kd^3\\
&+4(v-h)d^2u+(8/3)(h-v)ds_3+(8h+(8/3)v)d^4\\
&+hu^2+((2/3)v-2h)s_4-\sum_{j=2}^4F(t_j)-E_5.
\end{align*}
The coefficient of $s_3$ is between zero and $1/50$. The combined
quartic terms $hu^2+((2/3)v-2h)s_4-\sum F(t_j)$ are bounded below
by $-(11/100)u^2$. Thus
\[
\frac{D_4}{u}\ge\frac{11}{1000}-\frac{393}{500}\rho^2
 -\frac{167}{500}\rho^4-\frac1{50}\rho-\frac65\rho^3>\frac1{200}.
\]
At $\rho=1/20$ the remaining rational margin is
$230633/80000000>0$.

If $\delta\le1/500$, then $a_i^2>1/5$, hence $a_i+1/2>9/10$, and
\[
\rho^2\le\sum_i(a_i-1/2)^2\le\frac{100}{81}\delta
\le\frac1{405}<\frac1{400}.
\]
Since $\prod_i a_i\le1/16$ by AM--GM and
$\delta\le\sum_i(a_i^2-m^2)^2\le(9/4)\rho^2$,
\[
h(T_4)-h(S_a)\ge\frac{\rho^2}{100}\ge\frac{\delta}{225}.
\]
The entropy comparison again follows from nonnegativity of relative
entropy and support containment.

\section{Certified scalar integrations}\label{app:small-integrals}
We specify the finite computations used in Section~\ref{sec:small} and
the two preceding appendices. All polynomial coefficients, mesh endpoints,
and comparison thresholds are rational. Polynomial identities and final
algebraic margins are checked exactly. Transcendental evaluations use
192-bit Arb balls with outward rounding. An assertion $I<c$ means that
the upper endpoint of the computed interval for $I$ is strictly below $c$.

For $0\le k<n$ and $0\le s\le1$, the density of a sum of $n$ uniforms
on $[0,1]$ is
\[
 f_n(k+s)=\frac{P_{n,k}(s)}{(n-1)!},\qquad
 P_{n,k}(s)=\sum_{j=0}^k(-1)^j\binom nj(k+s-j)^{n-1}.
\]
Its values and derivatives are enclosed by Horner evaluation of these
integer polynomials. Symmetry reduces all integrals to $[0,n/2]$.

\paragraph{Gibbs slacks.}
For a twice differentiable function $g$ on an interval of length $\ell$
and midpoint $m$, we use
\[
 \left|\int g(x)\,dx-\ell g(m)\right|
 \le\frac{\ell^3}{24}\sup|g''|.
\]
For $Z_{W_n}$ we integrate $e^{-W_n}$ on $[0,8]$ and double the result.
There are $2^{15}$ equal cells for $n=3,4$ and $2^{17}$ for $n=5,6$.
On each complete cell the second derivative is enclosed using
\[
 (e^{-W})''=((W')^2-W'')e^{-W}.
\]
The displayed coefficients give $W'(8)>0$ and $W''(x)>0$ for $x\ge8$.
For the latter check, retain the leading term of $W''$ and its negative
terms, evaluate them at $8$, and multiply their positive sum by
$(x/8)^6$; this is a lower bound on $W''(x)$. Convexity then bounds both
omitted tails by $2e^{-W(8)}/W'(8)$.

To compute $h(X_*)$, use
\[
 h(X_*)=-\int_0^n f_n(t)\log f_n(t)\,dt+\tfrac12\log(12/n).
\]
On the outer interval $[0,1]$, $f_n(t)=t^{n-1}/(n-1)!$, so its entropy
contribution and its reflected copy are integrated exactly. Every
remaining cell lies in a region where $f_n$ is bounded away from zero.
Midpoint integration there uses the identity
\[
 (-f\log f)''=-f''(1+\log f)-(f')^2/f.
\]
The mesh length is $2^{-13}$ for $n=3,4$ and $2^{-15}$ for $n=5,6$.
The diagonal polynomial moments are rational, so these enclosures prove
the strict slack bounds in \eqref{eq:small-parameters}.

\paragraph{Local constants in dimensions three and four.}
All constants in Appendix~\ref{app:small-three} are elementary
expressions involving rational numbers, square roots, and logarithms;
they are enclosed directly. In Appendix~\ref{app:small-four}, the
integrands defining $J_2,J_3$, and the central entropy contribution are
enclosed uniformly on $2^{16}$ equal subintervals of $[0,1]$ and summed.
The outer entropy integral on $[1,2]$ is explicit. These interval sums
verify every numerical bound displayed in that appendix. The final
Taylor-remainder and neighborhood inequalities are rational checks.

\paragraph{The Hessian and averaged remainder.}
For \eqref{eq:small-kappa}, write $a=f_{n-2}(t-1)-f_n(t)$. The outer
interval is again integrated exactly. On each remaining interval,
midpoint quadrature with mesh length $2^{-12}$ uses
\[
 (a\log f)''=a''\log f+2a'f'/f
                         +a\bigl(f''/f-(f'/f)^2\bigr).
\]
This gives $\kappa_5>27/1000$ and $\kappa_6>17/1000$.

For $M_n(\gamma)$, the contribution of $0\le t\le1/2$ is exactly
\[
 \frac{4(n-1)}{(1-\gamma)^3n(n-1)!2^n}.
\]
On $[1/2,n/2]$, use intervals of length $2^{-11}$. For every such
interval $[a,b]$, enclose the density and $t^3$ on the full interval,
and bound the moving supremum on
$[(1-\gamma)a,\min(n/2,(1+\gamma)b)]$. Split this latter interval at
the spline knots and then into $32$ equal subintervals on each piece.
With $P=P_{n,k}$,
\[
 (\log P)'''=\frac{P^2P'''-3PP'P''+2(P')^3}{P^3}.
\]
The numerator is enclosed on each full subinterval; its absolute upper
endpoint is divided by the cube of a strictly positive lower endpoint
for $P$. Summing the resulting upper integrals proves
$M_5(29/500)<77/10$ and $M_6(1/25)<13/2$.
This procedure bounds every point of each integration interval, including
the full interval of the moving supremum.

The accompanying programs \texttt{certify\_small34.py} and
\texttt{certify\_small56.py}, with the shared routines in
\texttt{small\_integrals.py}, implement these computations and record the
resulting intervals and rational margins.
\section{A sharp inequality for the tilted variance}\label{app:langevin-paper}
Write $s=\sinh t$ and $c=\cosh t$. Direct differentiation gives
\[
 K'=\frac{tc-s}{ts},\qquad K''=\frac{s^2-t^2}{t^2s^2},\qquad
 -K'''=\frac{2(s^3-t^3c)}{t^3s^3}.
\]
All denominators are positive. The inequality of
Lemma~\ref{lem:langevin-paper} is therefore equivalent to $G(t)>0$, where
\[
 G(t)=10t^4s^4(s^3-t^3c)-54(tc-s)(s^2-t^2)^3.
\]
The following exact identity expands $32G$ in hyperbolic functions:
\begin{align*}
32G={}&(5t^4+27)\sinh7t-27t\cosh7t\\
&-(35t^4+324t^2+189)\sinh5t
  +(-20t^7+324t^3+135t)\cosh5t\\
&+(1401t^4+1620t^2+567)\sinh3t
  +(60t^7-1296t^5-972t^3-243t)\cosh3t\\
&-(1728t^6+4063t^4+3240t^2+945)\sinh t\\
&+(1688t^7+1296t^5+648t^3+135t)\cosh t.
\end{align*}
Only odd powers occur. Let $A_N=N![t^N](32G)$, and use
$(N)_k=N(N-1)\cdots(N-k+1)$. Exact expansion gives $A_N=0$ for odd
$N\le15$, while
\[
 A_{17}=2649777242112000,\qquad
 A_{19}=551423461860311040.
\]
For $N\ge17$, the contribution of the terms with argument $7t$ is
\[
 7^{N-4}[5(N)_4-9261N+64827]
 \ge\tfrac52(N)_4\,7^{N-4}=:L_N.
\]
Indeed, after putting $N=17+z$, the difference in brackets is
\[
 \tfrac52z^4+155z^3+\tfrac{7195}{2}z^2+27784z+50190>0.
\]
Discard every positive term with argument $5t,3t,t$, and let $B_N$ be
the sum of the absolute values of their negative coefficients, multiplied
by $N!$. Each summand is of the form $b_0(N)_k b^{N-k}$, with
$b\le5$ and $k\le7$. For odd $N\ge21$, its ratio to $L_N$ decreases
under $N\mapsto N+2$, because the ratio multiplier is at most
\[
 \frac{25}{49}\frac{(N-2)(N-3)}{(N-5)(N-6)}<1.
\]
At $N=21$, direct integer arithmetic gives
\[
 L_{21}=83537617572806033700,
 \qquad B_{21}=78199949030989113153.
\]
Thus $L_N>B_N$ for all remaining odd $N$, and all $A_N$ with $N\ge17$
are positive. The power series is entire, so $G(t)>0$ for every $t>0$.
Finally $K'=t/3+O(t^3)$, $K''=1/3+O(t^2)$, and
$-K'''=2t/15+O(t^3)$ give the sharp limit $54/5$.

\end{document}